\documentclass[aps,prd,preprint,superscriptaddress,showpacs,floatfix,nobibnotes]{revtex4-1}

\usepackage{latexsym}
\usepackage{amsmath}
\usepackage{amssymb}
\usepackage{graphicx}
\usepackage{longtable}

\usepackage{bm}
\usepackage{epsfig}
\usepackage{subfigure}
\usepackage{float}

\newcommand{\ket}[1]{\ensuremath{|#1\rangle}}			% ket
\newcommand{\iprod}[2]{\ensuremath{\langle#1|#2\rangle}}	% inner product
\newcommand{\bea}{\begin{eqnarray}}
\newcommand{\eea}{\end{eqnarray}}
\def\R{\mathbb R}    %%%%%%%%% the set of real numbers

\newcommand{\bpm}{\begin{pmatrix}}
\newcommand{\epm}{\end{pmatrix}}

\begin{document}

\title{On the geometric measure of entanglement for  pure states}

\author{M.E. Carrington}
\email[]{carrington@brandonu.ca} \affiliation{Department of Physics, Brandon University, Brandon, Manitoba, R7A 6A9 Canada}\affiliation{Winnipeg Institute for Theoretical Physics, Winnipeg, Manitoba}

\author{G.~Kunstatter}
\email[]{gkunstatter@uwinnipeg.ca} \affiliation{Department of Physics, University of Winnipeg, Winnipeg, Manitoba, R3B 2E9 Canada
}\affiliation{Winnipeg Institute for Theoretical Physics, Winnipeg, Manitoba}

\author{J.~Perron}
\email[]{jarradperron@hotmail.com} \affiliation{Department of Physics, Brandon University, Brandon, Manitoba, R7A 6A9 Canada}\affiliation{Winnipeg Institute for Theoretical Physics, Winnipeg, Manitoba}

\author{S.~Plosker}
\email[]{ploskers@brandonu.ca} \affiliation{Department of Mathematics and Computer
Science, Brandon University, Brandon, Manitoba, R7A 6A9 Canada}
\affiliation{Winnipeg Institute for Theoretical Physics, Winnipeg, Manitoba}

\begin{abstract}
The geometric measure of entanglement is  the distance or
angle between an entangled target state and the nearest unentangled  state. Often one considers the geometric measure of entanglement for highly symmetric entangled states because it simplifies the calculations and allows for analytic solutions. Although some symmetry is required in order to deal with large numbers of qubits, we are able to loosen significantly the restrictions on the highly symmetric states considered previously, and consider several generalizations of the coefficients of both target and unentangled states. This allows us to compute the geometric entanglement measure for larger and more relevant classes of states.
\end{abstract}

%\thanks{This research was supported by the Natural Sciences and Engineering Research Council (NSERC) of Canada and the Brandon University Research Committee (BURC) Research Award. }

\email{carrington@brandonu.ca, g.kunstatter@uwinnipeg.ca, perronjr15@brandonu.ca,  ploskers@brandonu.ca}

\maketitle

\section{Introduction}
Quantum entanglement is a key ingredient of quantum information theory and has been the subject of intense scrutiny for several years. Apart from its fundamental role in our understanding of deep conceptual issues in quantum mechanics in general and the quantum mechanical behaviour of black holes in particular, quantum entanglement also has more practical significance as a potential resource in quantum computing. The concept of entanglement is only useful, however, if it can be rigorously quantified in a large variety of different physical situations.

Several proposals for the definition of quantum entanglement exist in the literature. The focus of the present paper is geometrical entanglement, which measures the distance between a target entangled   pure state   $|\psi\rangle$ and the
closest separable state  $|\phi\rangle$: $D=|| |\phi\rangle-|\psi\rangle||$. This geometric measure of entanglement was first introduced in \cite{Shim95}. An equivalent approach is to measure the
angle $\theta$ between the  state $|\psi\rangle$ and the
closest separable state $|\phi\rangle$  \cite{WG03}. The closest separable state is not necessarily unique.

In \cite{Meg2010,Meg2011}, the authors considered the case where the closest separable state
was unnormalized rather than normalized, which simplifies calculations in some cases. It is possible to show in some generality \cite{Meg2010,Clement2012} that the normalized and unnormalized closest separable states found by these methods lie along the same ray in Hilbert space. The two methods therefore yield the same angle and the corresponding distance measures are related by a simple geometrical formula.  The choice of which of the two  entanglement measures to use therefore comes down to calculational convenience akin to a choice of parametrization. 

Given the existence of a variety of  entanglement measures it is important  to determine which of them are most relevant and useful to the situation at hand. To this end one must study the relationships between the various definitions. If two measures are equivalent in the sense that they always agree as to which of any two given states is more entangled, then in any particular context one can use the one that is easier to calculate. For example,  it is straightforward to derive analytically the relationship between quantum fidelity and the geometric measure of entanglement. 
 On the other hand, the relationship between two measures of entanglement is often more complicated:    two measures can disagree about which of two states is more entangled, and some measures of entanglement  often  give an inconclusive result (e.g.\ majorization). When an analytic relationship between two measures of entanglement is unknown or does not exist, one can only compare them by looking at specific examples of target states.  
Ultimately,  our goal is to study these issues by comparing implications of the geometric measure of entanglement with those of other measures.

 When dealing with complex systems with large numbers of qubits, solutions can only be obtained using certain simplifying assumptions. 
 Traditionally one restricts to target states with a high degree of symmetry. 
A number of papers have considered target states with a kind of permutation symmetry (such states are unchanged under permutation of any pair of qubits), and assumed that the closest separable state must exhibit the same symmetry.  
The GHZ, W, and  Dicke states are specific highly symmetric states that are frequently studied in the literature.  
The symmetry of these states allowed the authors of \cite{Meg2010,Meg2011} to find analytic solutions for their geometric measure of entanglement.
Several general results were found for geometric entanglement using normalized separable states: in \cite{Hay08} it was proven that at least one of the closest symmetric states mirrors the same symmetry as the target state; the special case in which  the target state has all non-negative coefficients was studied independently in \cite{Hay09, Wei10};
the fact that the closest separable state to a symmetric target  state  \emph{necessarily} has the same symmetry was proven in \cite{Hub09}. 

The purpose of the present paper is to relax previous restrictions and calculate  geometric entanglement for two new classes of symmetric states, each of which contains an arbitrarily large number of members. 
 Our results therefore make it possible to perform more  general comparisons of geometric entanglement with other measures. 
%We consider several generalizations of the coefficients of the target and unentangled states.
Specifically, we generalize \cite{Meg2010,Meg2011} in several significant ways. For one thing, we allow the coefficients of the separable states to be complex. The prior restriction to real coefficients was  not physically motivated; rather, it was used to simplify calculations. More significantly, we look at two types of generalized target states: First,
we consider a target state that is a linear combination of Dickie states with different
numbers of spin-up qubits. This provides a many-parameter family of target states for which the geometrical entanglement can be calculated in a straightforward way using a combination of analytic and numerical techniques. Second, we consider target states that are invariant under interchange of any pair of even qubits and/or odd qubits.  
 Our results generalize work done in \cite{Meg2010,Meg2011, WG03} and shed light on some of the  counterexamples  that arise in  \cite{Hub09}.

 The paper is organized as follows: the next Section establishes our notation, while Section III presents a detailed analysis  of the linear combinations of Dicke states. Section IV studies target states that are invariant under the interchange of a pair of even and/or odd qubits. Finally, Section V closes with some conclusions and prospects for further work.

\section{Notation}\label{sec:GHZW} 

We begin by defining our notation, following Refs. \cite{Meg2010,Meg2011}. We consider a multipartite system $\mathcal{H}=\mathcal{H}_A\otimes \mathcal{H}_B\otimes\mathcal{H}_C\otimes\cdots$ of $q$ qubits. The subsystems are labelled $A,B,C,\dots$ to represent the different parties Alice, Bob, Charlie, \dots The subsystems have dimension $u,v,w,\dots$ such that $n=u\cdot v\cdot w\cdots$. We fix an arbitrary set of basis states $|i\rangle$ for system $A$,  $|j\rangle$ for system $B$, $|k\rangle$ for system $C$, etc. Using this notation we write:
\bea
\label{A-def}
\ket{A}=\sum_{i=0}^{u-1} a_i\ket{i}\in \mathcal{H}_A,~~\ket{B}=\sum_{j=0}^{v-1} b_j\ket{j}\in \mathcal{H}_B,~~\ket{C}=\sum_{k=0}^{w-1} c_k\ket{k}\in \mathcal{H}_C,~~\dots
\eea
We consider an arbitrary normalized entangled pure state $|\psi\rangle$ written:
\bea
\label{psi}
\ket{\psi} = \sum_{i=0}^{u-1}\sum_{j=0}^{v-1}\sum_{k=0}^{w-1}\cdots \chi_{ijk\cdots} \ket{i}\otimes \ket{j}\otimes \ket{k}\cdots~,~~~\iprod{\psi}{\psi}=1\,.
\eea
We also consider a product state $\ket{\phi^\prime}$ which is not necessarily normalized and can be written:
\bea
\label{phi}
&& \ket{\phi^\prime} = \ket{A}\otimes\ket{B} \otimes \ket{C}\otimes\ldots
   = \sum_{i=0}^{u-1} a_i\ket{i} \otimes 
   \sum_{j=0}^{v-1} b_j\ket{j} \otimes
   \sum_{k=0}^{w-1} c_k\ket{k} \otimes\ldots \\
\label{norm}
&& \iprod{\phi^\prime}{\phi^\prime} = N_AN_BN_C\ldots; ~~
N_A =\iprod{A}{A}=\displaystyle \sum_{i = 0}^{u-1}a_i^*a_i,~~
N_B =\iprod{B}{B} = \displaystyle \sum_{j = 0}^{v-1}b_j^*b_j,~~\cdots
\eea
The distance between the states $\ket{\psi}$ and $\ket{\phi^\prime}$ can be written generally:
\bea
\label{dist}
D^2 &&= || |\phi^\prime\rangle-|\psi\rangle||^2=\iprod{\phi^\prime - \psi}{\phi^\prime - \psi} \,,\nonumber\\[2mm]
&& = 1 - \iprod{\phi^\prime}{\psi}-\iprod{\psi}{\phi^\prime}+N_A N_BN_C\ldots\nonumber\\[2mm]
&&
= \left( \sum_{i = 0}^{u-1}\sum_{j = 0}^{v-1}\sum_{k = 0}^{w-1}\cdots \right)
\left(a_i^*b_j^*c_k^*\ldots - \chi_{ijk\cdots}^*\right)
\left(a_ib_jc_k\ldots - \chi_{ijk\cdots}\right)\,.
\eea

In the remainder of this paper we consider a system of $q$ spin 1/2 qubits, which means $u=v=w\cdots=2$ in equations (\ref{A-def}-\ref{dist}). 
We consider only real target states, but the coefficients of the separable state can be complex.  This differs from some of the examples in \cite{Hub09} where the authors take the target state $\ket{\psi}$ to be real and consider minimizing $D$ over all  real product vectors. The target state has the general form (see (\ref{psi})),
\bea
\label{psi_qubit}
\ket{\psi} = \sum_{i=0}^{1}\sum_{j=0}^{1}\sum_{k=0}^{1}\cdots \chi_{ijk\cdots} \ket{i}\otimes \ket{j}\otimes \ket{k}\cdots~,~~~\iprod{\psi}{\psi}=1\,.
\eea

The components of the product state in (\ref{phi}) can be written
\bea
\label{phi2}
|A\rangle = e^{i\Theta_a}| a\rangle\,, ~~
|a\rangle = \bpm a_1\\ e^{i\theta_a} a_2\epm\, \textnormal{ and similarly for } | B\rangle, | C\rangle, \dots
\eea
%
%For a system of $q$ spin 1/2 qubits, the target state has the general form,
%\bea
%\label{psi_qubit}
%\ket{\psi} = \sum_{i=0}^{1}\sum_{j=0}^{1}\sum_{k=0}^{1}\cdots \chi_{ijk\cdots} %\ket{i}\otimes \ket{j}\otimes \ket{k}\cdots~,~~~\iprod{\psi}{\psi}=1\,.
%\eea
In \cite{Meg2010,Meg2011} the authors considered target states that are symmetric under the interchange of any two qubits. The most general example they studied was the $q$-qubit spin 1/2 Dicke state which contains all possible combinations
of $p$ entries of ``1'' and $q-p$ entries of ``0'':
\bea
\label{dickie-state}
|D_p\rangle=\frac1{\sqrt{{q\choose p}}}\sum_{\ell}P_\ell\{|1\rangle^{\otimes p}\otimes|0\rangle^{\otimes{q-p}}\},
\eea
where $\sum_{\ell}P_\ell\{\cdot\}$ denotes the sum over all possible permutations.
The W state is the $q$-qubit state of the form
\bea
\label{Wstate}
    |W\rangle = \frac{1}{\sqrt{q}}(|100...0\rangle + |010...0\rangle + ... + |00...01\rangle),
\eea
which corresponds to a Dicke state with $p=1$.
%

%One can take advantage of the symmetry of these states which greatly simplifies the calculations of the geometric measure of entanglement. However, for practical purposes we wish to find solutions to a larger family of generalizations of these states, rather than constraining ourselves to only these exact states. Our analysis uses linear combinations of these symmetric states, which breaks down some of the symmetry and makes the calculations not as straightforward as in earlier works. 

\section{Linear Combinations of Dickie States}

\subsection{Distance Measure}
We consider a target state that is a linear combination of Dickie states with different numbers of spin-up qubits
\bea
\label{dickie2}
|\psi\rangle = \sum_{p=0}^q f_p|D_p\rangle \quad\textnormal{ where } f_p\in \R, \sum f_p^2=1.
\eea
Substituting (\ref{phi}), (\ref{phi2}) and (\ref{dickie2}) into the distance measure (\ref{dist}) with $\tilde a_1 = a_1$ and $\tilde a_2 = e^{i\theta a}a_2$, we obtain 
\bea
D^2 = 1+N_a N_b N_c\cdots - \langle\psi|\phi'\rangle - \langle\phi'|\psi\rangle\,,\\
\langle\psi|\phi'\rangle = e^{i(\Theta_a+\Theta_b+\cdots)}\chi_{ijk\cdots} \tilde a_i \tilde b_j \tilde c_k \cdots
\eea
We define
\bea
Z&=& e^{i(\Theta_a+\Theta_b+\cdots)}\,,~~~z_a = e^{i\theta_a}\,,~~~z_b = e^{i\theta_b} \cdots \textnormal{ so that }\\
|\phi'\rangle &=& Z|\phi\rangle \textnormal{ where } |\phi\rangle = | a\rangle \otimes | b\rangle \otimes |c\rangle \cdots
\eea
and the distance measure becomes
\bea
D^2 = 1+N_a N_b N_c\cdots - Z \langle\psi|\phi\rangle - \frac{1}{Z} \langle\phi|\psi\rangle\,.
\eea
We will consider real target states that are symmetric under the interchange of any two qubits, and we use the ansatz that the product state has the same symmetry, which means
\bea
&& a_1=b_1=c_1 \cdots\,,~~N_a=N_b=N_c \cdots \equiv N\\
&& \theta_a = \theta_b=\theta_c \cdots  ~~\text{or}~~ z_a=z_b=z_c \cdots\equiv z\,.
\eea
When we contract $|D_p\rangle$ with the product state we obtain
\bea
\langle D_p|\phi\rangle = {q\choose p} \,\frac1{\sqrt{{q\choose p}}} a_1^{q-p} (a_2 z)^p\,,
\eea
where the first combinatoric factor comes from the number of non-zero terms in the contraction, and the second is the normalization of the Dickie state. Absorbing all the combinatoric factors into the definition of the $f$'s, we can write
\bea
\langle \psi|\phi\rangle  = \sum_p f_p  a_1^{q-p} (a_2 z)^p\,.
\eea

We now rewrite the variables $\{a_1,a_2\}$ in terms of two different variables $\{N,r\}$ using
\bea
\label{Nandr}
a_2 = r a_1\,,~~a_1 = \sqrt{\frac{N}{1+r^2}}\,.
\eea
Since either $a_1$ or $a_2$ could be zero, we will consider the cases $r=0$ and $r\rightarrow \infty$ separately in section \ref{section:specialcases}. We take $\theta$ and $\Theta$ from 0 to $2\pi$ and therefore we can assume without loss of generality that $a_1$, $a_2$ and $r$ are positive. 
Using this notation we have
\bea
a_1^{q-p}a_2^p = \left(\frac{\sqrt{N}}{\sqrt{1+r^2}}\right)^q r^p\,.
\eea

We  define real and imaginary combinations of the   variables $\{z,Z\}$ as, 
\bea
Z_{\Re}(p) &=& Z z^p+\frac{1}{Z z^p}= 2\cos(\Theta + p \theta) \\
Z_{\Im}(p) &=& -i(Z z^p-\frac{1}{Z z^p})=2\sin(\Theta + p\theta)\,,
\eea
where $\Theta = \sum (\Theta_a + \Theta_b+\cdots)$.
We further write the real quantities 
\bea
\label{gens}
g_{\Re}(q,m) &=& \sum_{p=0}^q p^m f_p Z_{\Re}(p) r^p =2 \sum_{p=0}^q p^m f_p \cos(\Theta + p \theta) r^p\\
\label{genI} g_{\Im}(q,m) &=& \sum_{p=0}^q p^m f_p Z_{\Im}(p) r^p = 2 \sum_{p=0}^q p^m f_p \sin(\Theta + p\theta) r^p\,,
\eea
where $m$ is taken to be a non-negative integer, 
and we define $p^m|_{m=0} = 1$ for all $p$ (including $p=0$). 

Using this notation, the distance measure is
\bea
\label{dsqN}
D^2 =1+N^q -\left[\frac{N}{1+r^2}\right]^{q/2} g_{\Re}(q,0)\,.
\eea
We want to minimize this distance. To find extrema, we take derivatives with respect to $\{N,r,\theta,\Theta\}$ and solve the four equations simultaneously. To identify the minima, we look at the Hessian. 

\subsection{Extremal Equations}

We use the relations:
\bea
\frac{\partial g_{\Re}(q,m)}{\partial \Theta} &=& -g_{\Im}(q,m)\\
\frac{\partial g_{\Im}(q,m)}{\partial \Theta}&=&  g_{\Re}(q,m)\\
\frac{\partial g_{\Re}(q,m)}{\partial \theta}&=& - g_{\Im}(q,m+1)\\
\frac{\partial g_{\Im}(q,m)}{\partial \theta}&=&  g_{\Re}(q,m+1)\\
\frac{\partial g_{\Re}(q,m)}{\partial r}&=& -\frac{1}{r} g_{\Im}(q,m+1)\\
\frac{\partial g_{\Im}(q,m)}{\partial r}&=& \frac{1}{r} g_{\Re}(q,m+1)\,.
\eea
This gives the following equations which determine the location of the extrema:
\bea
\label{Neqn1}
\frac{\partial D^2}{\partial N}&=&q\frac{N^{q/2-1}}{(1+r^2)^{q/2}}\left(\big[N (1+r^2)\big]^{q/2}-\frac{1}{2}g_{\Re}(q,0)\right) =0\\
\label{reqn1}
\frac{\partial D^2}{\partial r}&=&\frac{N^{q/2}}{(1+r^2)^{q/2+1}}\left(q r g_{\Re}(q,0)-\frac{1+r^2}{r}g_{\Re}(q,1)\right) =0\\
\label{zeqn1}
\frac{\partial D^2}{\partial\theta}&=&\frac{N^{q/2}}{(1+r^2)^{q/2}}g_{\Im}(q,1) =0\\
\label{Zeqn1}
\frac{\partial D^2}{\partial\Theta}&=&\frac{N^{q/2}}{(1+r^2)^{q/2}}g_{\Im}(q,0) \,.
\eea

The solution $N=0$ always exists, but corresponds to a maximum, so we consider $N>0$. Both $N$ and $r$ are always real, and   we can therefore remove an overall factor of $\frac{N^{q/2}}{(1+r^2)^{q/2}}$  from (\ref{Neqn1} - \ref{Zeqn1}) since it cannot affect the solutions (the case $r\to\infty$ is discussed in section \ref{section:specialcases}).
We rewrite the resulting equations:
\bea
\label{Neqn}
g_{\Re}(q,0)&=&2\big[N (1+r^2)\big]^{q/2} \\
\label{reqn}
\frac{1}{r} g_{\Re}(q,1) &=& \frac{q r}{1+r^2} g_R(q,0) = \frac{2 qr}{1+r^2}\big[N (1+r^2)\big]^{q/2}  \\
\label{zeqn}
g_{\Im}(q,1)&=&0 \\
\label{Zeqn}
g_{\Im}(q,0)&=&0 \,.
\eea
Note that $\lim_{r\to 0}\frac{1}{r} g_{\Re}(q,1) \sim r^0$. 

From (\ref{genI}) it is clear that (\ref{zeqn}, \ref{Zeqn}) give $z, Z\in \{\pm 1\}$ if the number of non-zero $f_p$'s is less than or equal to two, which justifies the ansatz of \cite{Meg2010,Meg2011}.
Substituting (\ref{Neqn}) into (\ref{dsqN}) we find that the minimal solution is:
\bea
\label{Dsqcrit}
D^2_{\text{c}} = 1-N^q\,,
\eea
which also agrees with the result of \cite{Meg2010,Meg2011}.

\subsection{Comparison of normalized and unnormalized distances}

One can explicitly verify the equivalence between using normalized and unnormalized separable states to define the geometrical entanglement. From (\ref{dist}) the distance measures for unnormalized and normalized product states can be written
\bea
\label{distances}
&& (D^2_c)_{\rm unnormalized} = 1-\cos^2\theta_c\\
&& (D^2_c)_{\rm normalized} = 2(1-\cos\theta_c)\nonumber
\eea
where $\theta_c$ is defined as 
\bea
\label{crit-angle}
\cos\theta_c:=\bigg|\frac{\langle \psi|\bar\phi\rangle}{\sqrt{\langle\bar\phi|\bar\phi\rangle  \langle\phi|\phi\rangle}}\bigg|\,.
\eea
Equations (\ref{Neqn1}-\ref{Zeqn1}) give 
\bea
\cos\theta_c =\frac{1}{2}\, \frac{1}{(1+r^2)^{q/2}} g_C(q,0)
\eea
which is independent of $N$. 
The three equations from the $r$, $z$ and $Z$ derivatives (\ref{reqn1}-\ref{Zeqn1}) have an overall factor 
\bea
F = \left(\frac{N}{1+r^2}\right)^{q/2}\,.
\eea
Assuming for the moment that $F\ne 0$ we can remove this factor and obtain three equations that are independent of $N$. 
Solving for $\{r,z,Z\}$ gives some solution $\{r_0,z_0,Z_0\}$, which gives from (\ref{crit-angle}) a critical angle, which gives from (\ref{distances}) two different distances with a definite relation between them. The conclusion, for $F\ne 0$, is that the geometric entanglement measures using either unnormalized or normalized product states, are equivalent. \\

Now we discuss the possibility that the factor $F=0$. 
If $N=0$ the unnormalized distance is maximum, so this cannot correspond to the closest state. In section \ref{section:specialcases} we consider the special cases $r=0$ and $r\to\infty$.
\subsection{The Hessian}
We can use the Hessian to determine if the extremal solutions are minima or maxima. The Hessian is a $2q\times 2q$ matrix of the second derivatives of the distance functions with respect to the parameters $\{N,r,z,Z \dots\}$, evaluated at a given extremum. For the extremum to be a local minimum, all eigenvalues of the Hessian must be positive, apart from the zero eigenvalues associated with the symmetries of the system. 
The Hessian cannot determine if a given solution is a local or global minimum. The Hessian for our situation is given below, with the order of components $\{N,r,\theta,\Theta\}$, and the definition $R=1+r^2$.
\bea
H = \left(
\begin{array}{cccc}
 A & 0 & 0 & 0 \\
 0 & B & X & Y \\
 0 & X & C & W \\
 0 & Y & W & D \\
\end{array}
\right)
\eea
\bea
A &=& \frac{q^2}{2} N^{q/2-2}R^{q/2} \\
r^2 B &=& \frac{1}{R} \left[q r^2 g_{\Re}(q,0)+\left((q-1) r^2+1\right) g_{\Re}(q,1)-R
   g_{\Re}(q,2)\right]\nonumber\\
  &=& \frac{2qr^2}{R^2} (NR)^{q/2}(2+qr^2) - g_{\Re}(q,2)\\
C &=&  g_{\Re}(q,2) \\
D &=& g_{\Re}(q,0)=2(NR)^{q/2}\\
r X &=& g_{\Im}(q,2) \\
Y &=& 0\\
W &=&  g_{\Re}(q,1) =\frac{2qr^2}{R}(NR)^{q/2}\,.
\eea  
The matrix is sufficiently sparce that we can find all the eigenvalues explicitly.
The characteristic equation is:
\bea
0= (A-\lambda)(B-\lambda)\left[
(C-\lambda)(D-\lambda)-X(D-\lambda)-W^2
\right]
\eea
which gives:
\bea
\lambda=A\,,~~\lambda=B\,,~~
\lambda = \frac{C+D-X}{2}\pm \frac{\sqrt{(C-D-X)^2+4W^2}}{2}\,.
\eea
The last two eigenvalues are positive if $D(C-X)>W^2$.

%%%%%%%%%%%%%%%%%%%%%%%%%%%%%%%%%%%%%%%%%%%%%%%%%%%%%%%%

\subsection{Special cases: $r=0$ and $r\to\infty$}
\label{section:specialcases}

These values of $r$ are at the edge of the parameter space so they could give a smallest distance which is not extremal. However, we can find the corresponding  distances analytically using (\ref{dsqN}).
For $r=0$ the distance is:
\bea
\label{D0soln}
D^2(0) = 1+N^{q}-2N^{q/2}f_0 \cos\Theta\,.
\eea
We cannot choose $f_0=0$ because this would give $D(0)^2=2=(D_c)_{\rm max}$ for normalized states  and $D(0)^2=1+N^q>[1=(D_c)_{\rm max}]$ for unnormalized states.
Eq. (\ref{D0soln}) shows that the minimum $D(0)$ comes from $\cos\Theta=\pm 1$ (independently of whether this distance is extremal).
For unnormalized states extremizing in the remaining variable $N$ gives $D(0)^2 = 1-f_0^2$.
For normalized states we can just set $N=1$ and get $D(0)^2 = 2(1-|f_0|)$.
%The Hessian shows that there is a zero eigenvalue that corresponds to rotation in $\theta$, which reflects the fact that the distance is independent of $z$ when $r=0$. 
%

For $r\to\infty$ the distance is:
\bea
\label{Dinsoln}
D(\infty)^2 = 1+N^{q}-2N^{q/2}f_q \cos\Theta\,.
\eea
In the same way as for the $r=0$ case we obtain the following results.
We cannot have $f_q=0$. 
The minimum $D(\infty)$ comes from $\cos\Theta=\pm 1$ (independently of whether this distance is extremal). 
For unnormalized states extremizing gives $D(\infty)^2 = 1-f_q^2$.
For normalized states we set $N=1$ and get $D(\infty)^2 = 2(1-|f_q|)$.
%The Hessian has a zero eigenvalue that corresponds to rotation in $\theta$, and an additional zero eigenvalue with eigenvector pointing in the $r$-direction. 

\subsection{Numerical testing}

The extremal equations (\ref{Neqn1}-\ref{Zeqn1}) cannot be solved analytically except in special cases, but we can solve them numerically. 
We consider only target states with three or more non-zero $f_i$'s where complex solutions can exist.
We used $q=4$ qubits and considered random values of $f_0, \dots, f_4$, which we write as a vector $\vec f = (f_0,f_1,f_2,f_3,f_4)$ for notational simplicity (we call this vector an $f$-vector below). 
Using 18450 different states, we solved the extremal equations to find the coefficients of the product state that extremizes the distance measure. We then  calculated the eigenvalues of the Hessian to identify the minimal solution. 
Solutions for which all eigenvalues are non-negative correspond to minima, provided that zero eigenvalues are associated with a symmetry of the distance measure. 
We found:
\begin{enumerate}
\item 17476 real solutions with positive eigenvalues

\item 39 real solutions with two positive eigenvalues and one zero eigenvalue 

\item 648 complex solutions with positive eigenvalues

\item 152 solutions at $r=0$, which agree with the corresponding  analytic solution (\ref{D0soln})

\item 135 solutions at $r\to\infty$, which agree with the corresponding  analytic solution (\ref{Dinsoln})

\end{enumerate}

Complex solutions occur rarely, and only when some components of $\vec f$ are negative.
States with real solutions and a zero eigenvalue have a lot of symmetry. 
Two examples are $\vec f=(1,1,1,1,1)$ and $\vec f=(-2,3,-1,3,-2)$. In all cases for which there is a zero eigenvalue, the minimal solution is $r=1$ and $z,Z\in \{\pm1\}$, and the eigenvector that corresponds to the zero eigenvalue points in the $r$ direction. 
%In this way, any zero eigenvalues are associated with the symmetries of the system, and so they do not pose a problem. 

\begin{figure}[H]
\begin{center}
\includegraphics[width=10cm]{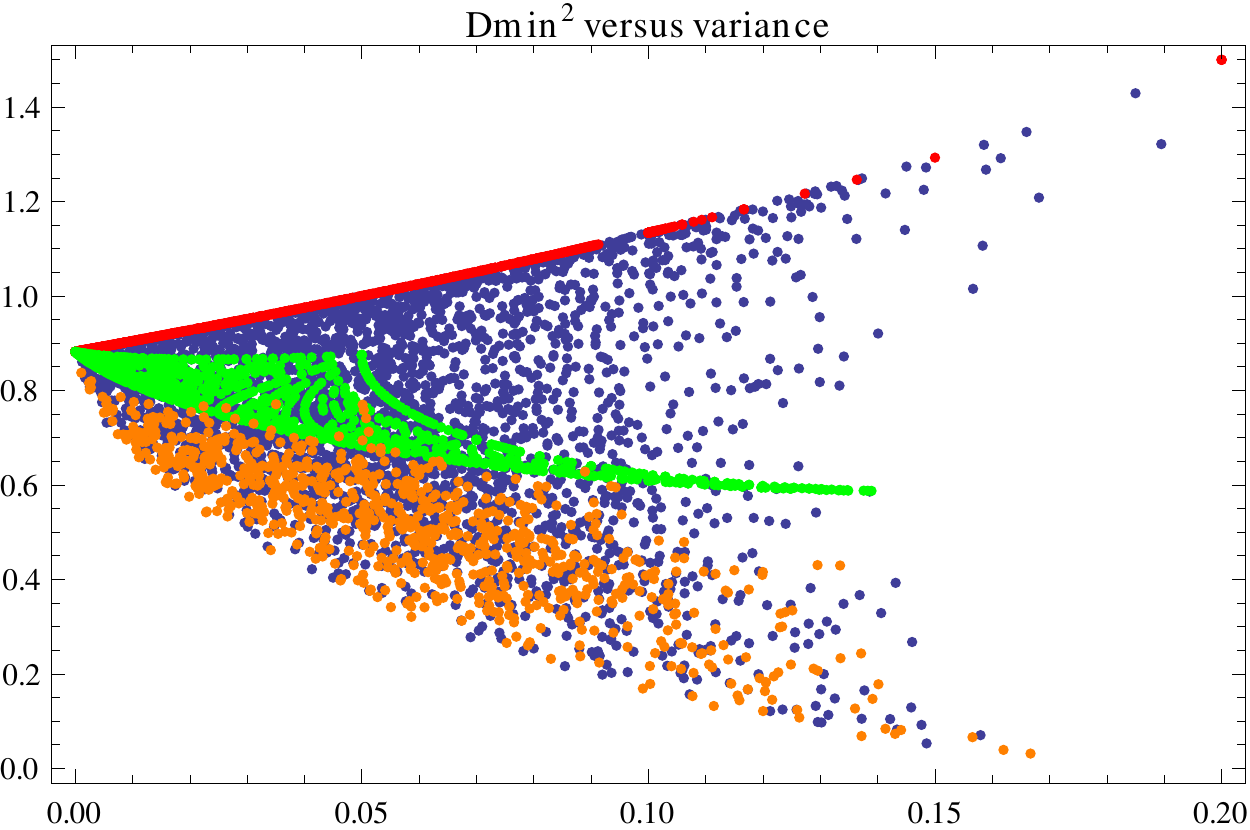}
\end{center}
\caption{The minimal distance as a function of the variance of the $f$-vector for $q=4$. The red dots represent the Gaussian-like vectors peaked at $p=2$.
The red point at the upper right is the $p=2$ Dickie state $\vec f = (0,0,1,0,0)$  and the red point at the far left is the vector $\vec f = (1,1,1,1,1)$. The green dots correspond to symmetric distributions peaked at both ends, while the orange dots are peaked at one end.\label{var-fig2}}
\end{figure}

We now study the  correlation between entanglement and the distribution of $f_p$ values. 
We define the variance of an $f$-vector 
\bea
\operatorname{Var}(\vec{f})=
  \frac{1}{q}\sum_{p=0}^q\big(f_p -E(\vec{f}\,)\big)^2 \,,~~~E(\vec{f})=\frac1{q+1}\sum_{p=0}^qf_p\,.
\eea
Dickie states correspond to $f$-vectors with one non-zero component. 
We consider $f$-vectors with only positive components and try to see if there is a correlation between the variance of the $f$-vector and the entanglement (quantified by the minimal distance) of the corresponding target state. Fig \ref{var-fig2}. shows a plot of the minimal distance against the variance of the $f$-vector. 
The behaviour of the data points seems random, except for the fact that all the values occur within a wedge that broadens towards increasing variance. 
In order to get more information, we consider separately three different specialized types of $f$-vectors which are represented with different coloured data points in the figure, as explained below.

\begin{enumerate}
\item Gaussian-like distributions centered around $p=2$ such that $f_p$ is largest at $p=2$ and smallest at  $f_0$ and $f_4$ (a graph of these $f$-vectors looks is bell-shaped). An example is $f = (0,1,2,1,0)$. The results obtained from this class of $f$-vectors are shown as red dots in Fig. \ref{var-fig2}. 

\item Inverted Gaussians with minimum at $p=2$ such that $f_p$ is smallest at $p=2$ and largest at  $f_0$ and $f_4$ (a graph of the $f$-vector has an inverse bell-shape). An example is $f = (4,1,0,1,4)$. The results are shown as green dots in Fig. \ref{var-fig2}. 

\item Distributions that correspond to the left or right side of a symmetric Gaussian. Examples are $\vec f = (4,4,3,2,1)$ or $\vec f = (0,1,1,2,4)$. The results are the orange dots in Fig. \ref{var-fig2}. 

\end{enumerate}

\begin{figure}[H]
\begin{center}
\includegraphics[width=10cm]{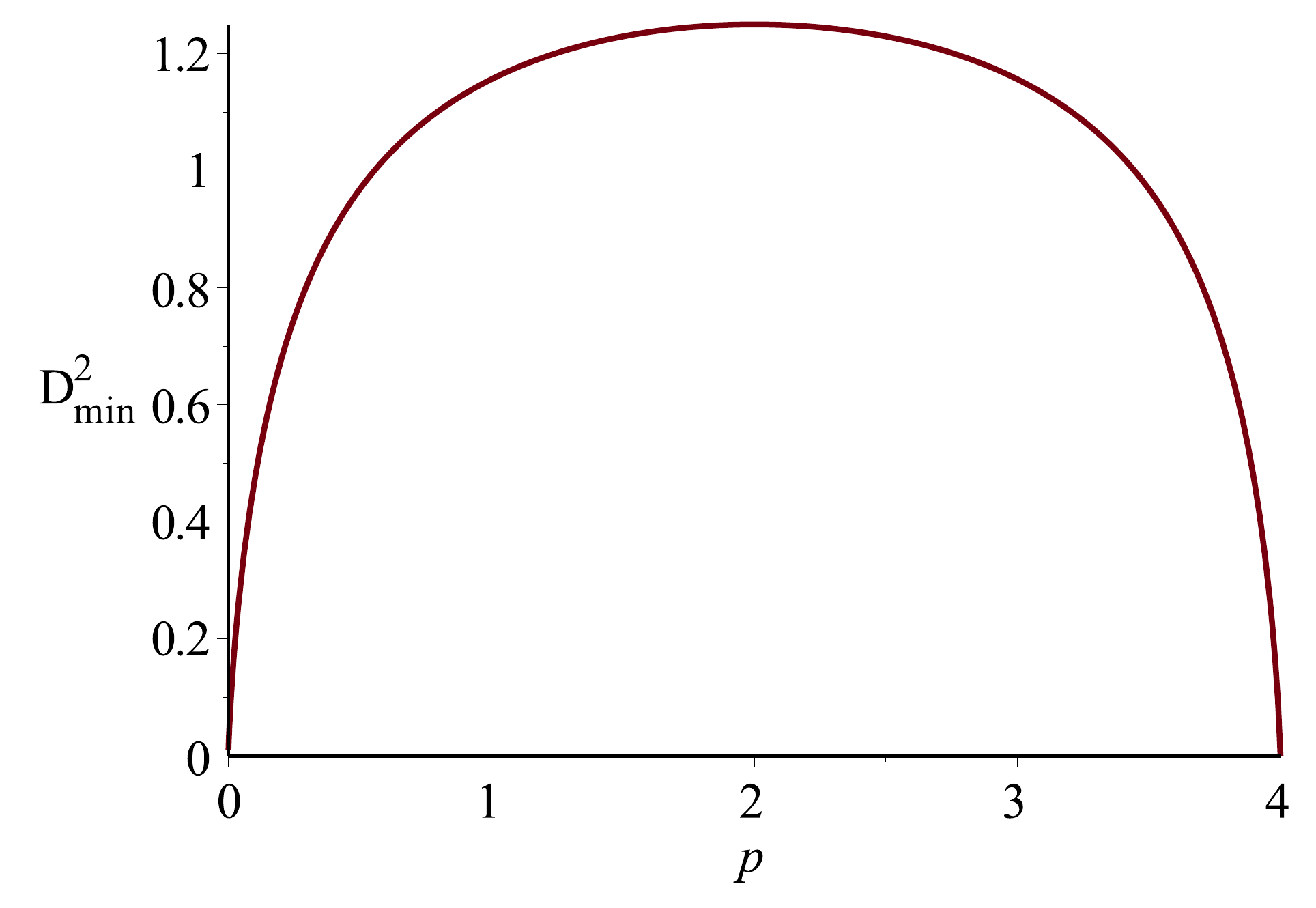}
\end{center}
\caption{Geometrical entanglement (normalized)  of pure $q=4$ Dicke states as a function of the number $p$ of ``1 entries.''  \label{Dicke}\label{fig:dicke}}
\end{figure}

The red point at the upper right of the figure is the $p=2$ Dickie  state which corresponds to $\vec f = (0,0,1,0,0)$.
This is the most entangled of the red states, and as shown in Fig. \ref{fig:dicke} the $p=2$ Dickie state is the most entangled of the Dickie states \cite{Meg2010}.
The red point at the far left (the least entangled red state) is the vector $\vec f=(1,1,1,1,1)$. 
The result is therefore that, as expected, adding linear combinations of other $p$'s to the most entangled Dickie state decreases entanglement. 
We also note that for a given variance, the red state always gives the maximum entanglement (the red line is the upper boundary for the wedge in Fig. \ref{var-fig2}).

The green dots also form a pattern, indicating a correlation between the entanglement and the coefficient of the $p=2$ Dickie state in the linear combination, but the plot is more complicated in this case. We speculate on the reason. In general adding more of the $p=2$ state (increasing the magnitude of $f_2$) would tend to increase the entanglement. In addition, one would expect that a larger variance in the $f$-vector would also have this effect by moving the target state further from the space of product states. For the red line these two effects (adding more $p=2$ and increasing the variance) combine constructively, while in the green data, the variance increases as more $p=2$ state is added, so the effects cancel to some extent. There is however additional interesting structure in the distribution that we do not as yet understand.

For the orange states the general trend is similar to the green states: as the Gaussian spreads and the variance decreases, one gets more contribution from the $p=2$ Dickie state. The former tends to decrease the entanglement while the latter tends to increase it. The net effect is more random than for the inverted Gaussian (green states).

\section{Symmetry of Even (Odd) Qubits}\label{sec:evenodd} 

%Although we have an inequality relating the entanglement of the GHZ-like (W-like) states to the GHZ (W) state, this gives no information describing the closest product state to a GHZ-like (W-like) state. Moreover, the aforementioned results \cite{Hay08, Hub09} do not apply to  the arbitrary coefficients case. Still, we would like to carry through calculations for exact solution when the target state has arbitrary coefficients, rather than all equal coefficients. We take steps in this direction in the examples below.

In this section, we consider target states that are invariant under interchange of any pair of even qubits and/or odd qubits. That is, any pair of Alice, Charleen, Ellen, Gertrude,... can interchange qubits, and any pair of Bob, David, Fred, Harry,... can interchange qubits, without changing the target state. We assume that the closest product state has the same symmetry. 
The product state has the form
\bea
\label{phieo}
&& \ket{\phi} = \ket{A}\otimes\ket{B} \otimes \ket{A}\otimes\ket{B}\otimes\ldots
\\
&& |A\rangle = e^{i\Theta_a}| a\rangle\,, ~~
|a\rangle = \bpm a_1\\ e^{i\theta_a} a_2\epm\,,  ~~
|B\rangle = e^{i\Theta_b}| b\rangle\,, ~~
|b\rangle = \bpm b_1\\ e^{i\theta_b} b_2\epm\,.
\eea

We comment that a notion of translationally invariant
states was introduced in \cite{Hub09} where 
the authors define a translationally invariant state as one which satisfies $\chi_{ijk\cdots}=\chi_{jk\cdots i}=\chi_{k\cdots ij}=\cdots$. 
This type of symmetry is not interesting for our calculations, because the closest product state to a translationally invariant target state, is not necessarily translationally invariant. An example is the target state $\ket{\psi}=\frac1{\sqrt2}(\ket{0101}+\ket{1010})$ which is translationally invariant. The closest separable states are the non-translationally invariant states $\ket{0101}$ and $\ket{1010}$. Notice however that both the target and product state are symmetric under interchange of even (odd) qubits. 

\begin{figure}[htb]
\begin{center}
\includegraphics[width=10cm]{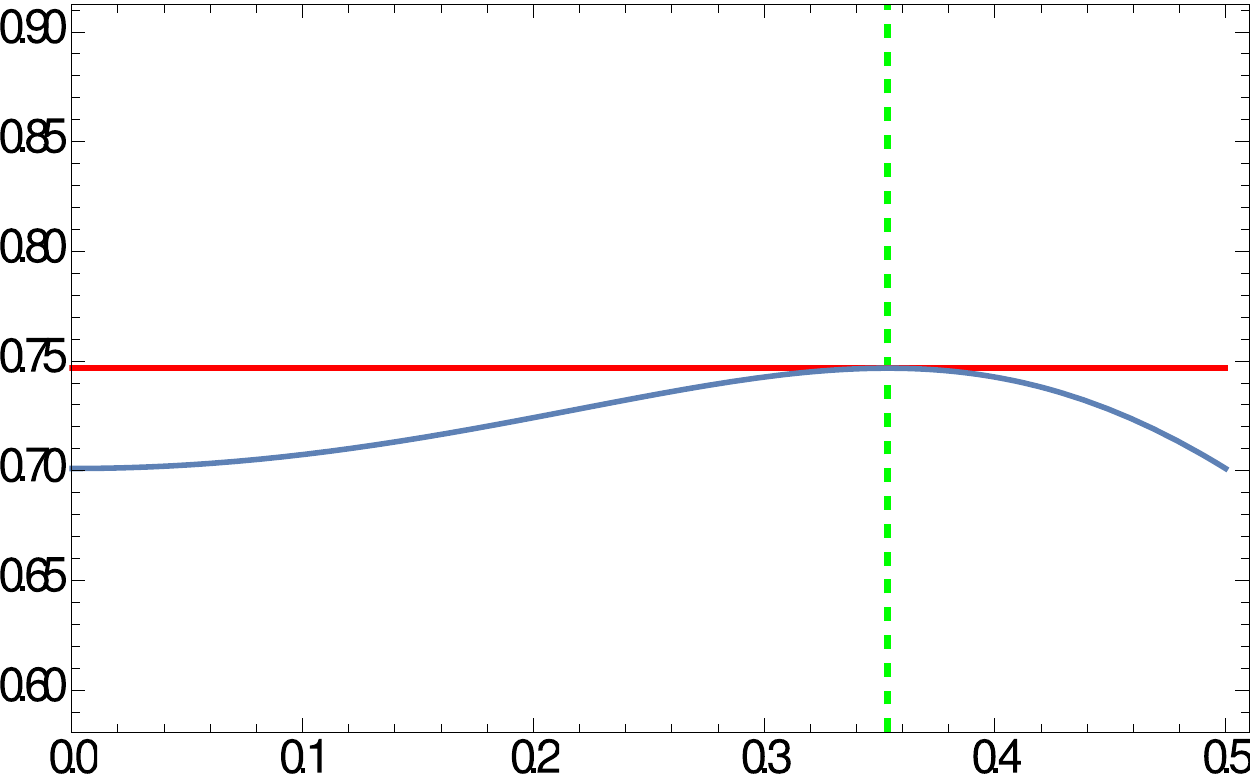}
\end{center}
\caption{The minimal distance squared as a function of $f$ for $q=4$. The horizontal red line is the W state and the vertical green line is the point where $f=m=\sqrt{2/q}$. \label{eo-fig}}
\end{figure}

As in equation (\ref{Nandr}), we define 
\bea
\label{Nandreo}
a_2 = r_a a_1\,,~~a_1 = \sqrt{\frac{N_a}{1+r_a^2}}\,,~~b_2 = r_b b_1\,,~~b_1 = \sqrt{\frac{N_b}{1+r_b^2}}\,.
\eea
We consider a `W-like' state of the form (see (\ref{Wstate}))
\bea
\label{Wlikestate}
    |W\rangle = f|100...0\rangle + m|010...0\rangle + f|0010...0\rangle + m|00010...0\rangle  + ... + m|00...01\rangle),
\eea
with $\frac{q}{2}(f^2+m^2)=1$. 
Here we assume the number of qubits $q$ is even. This restriction is not important when $q\rightarrow \infty$. Substituting into the distance measure we obtain:
\bea
&& D^2 = 1+ N_a^{q/2} N_b^{q/2} +Z C+\frac{1}{Z} C^*\\
&&C= \frac{1}{2} q \left(\frac{N_a}{r_a^2+1}\right)^{q/4}
   \left(\frac{N_b}{r_b^2+1}\right)^{q/4} \left(f z_a r_a+m z_b r_b\right)\,.
\eea
Solving the $Z$ equation (\ref{Zeqn1}) immediately we find that the distance can be written
\bea
D^2 &&= 1+N_a^{q/2} N_b^{q/2} - \sqrt{C\,C^*}\,.
\eea
Minimizing the distance is equivalent to maximizing the quantity in the square root, which can be separated into two pieces, one of which does not depend on $z_a$ or $z_b$. 
Writing the result in terms of the angles $\theta_a$ and $\theta_b$ we obtain
\bea
&& C = P_1 P_2 \\
&& P_1 = \frac{1}{4} q^2 \left(\frac{N_a}{r_a^2+1}\right){}^{q/4}
   \left(\frac{N_b}{r_b^2+1}\right){}^{q/4} \\
&& P_2 = 2 f m r_a r_b \cos \left(\theta _a-\theta _b\right)+f^2 r_a^2+m^2 r_b^2\,.
\eea
We see that $C$ is maximized when the cosine equals plus or minus one:
\bea
\label{cos-soln}
\cos \left(\theta _a-\theta _b\right) = \text{sgn}[fm]\,.
\eea
Once again, the extremal equations from the derivative of the variables which give the normalization of the product state qubits ($N_a$ and $N_b$) are decoupled from the other extremal equations, and therefore we work from this point on with normalized product states. Setting $N_a=N_b=1$ and differentiating with respect to $r_a$ and $r_b$ the expression for $C$ obtained using (\ref{cos-soln}),
 we obtain:
\bea
&& F = \frac{4 \left(r_a^2+1\right)^{q/2} \left(r_b^2+1\right)^{q/2}}{q^2 \left(f
   r_a+m \text{sgn}[fm] r_b\right)}\\[2mm]
&&   F \left(r_a^2+1\right) \cdot m \big[ (2 f m-m^2 q r_a r_b \text{sgn}[fm] -(q-2)fm r_a^2) \big]= 0\\[2mm]
&&   F \left(r_b^2+1\right) \cdot f \text{sgn}[fm]\big[(2 f m-f^2 q r_a r_b \text{sgn}[fm] -(q-2)fm r_b^2) \big]= 0\,.
\eea
Solving these equations gives
\bea
&& r_a(f,m)=\frac{q \left(\sqrt{f^4 q^2+2 f^2 m^2 ((q-8) q+8)+m^4 q^2}-m^2 q\right)-f^2 ((q-8)
   q+8)}{4 f^2 (q-2) (q-1)}\nonumber\\[2mm]
&& r_b(f,m)= r_a(m,f)
\eea
and the critical distance is 
\bea
D_{min} = 2-q \left(\frac{1}{r_a^2+1}\right)^{q/4} \left(\frac{1}{r_b^2+1}\right)^{q/4}\sqrt{
   \left(2 |f \,m | r_a r_b+f^2 r_a^2+m^2 r_b^2\right)}\bigg|_{r_a=r_a(f,m)\,,~r_b=r_b(f,m)}\nonumber\\
   \eea
where the parameters $f$ and $m$ can be anything that satisfy the normalization condition $\frac{q}{2}(f^2+m^2)=1$. 
When $f=m=\sqrt{2/q}$ we recover the result for the W state \cite{Meg2011}
\bea
D_{min} = 2-2 \left(\frac{q-1}{q}\right)^{\frac{q-1}{2}}\,\,.
\eea
In Fig. \ref{eo-fig} we show the  minimal distance squared as a function of $f$. The most entangled state is the W state. 

\section{Conclusions}

We have calculated the geometrical entanglement for two classes of multi-partite states that have not been previously treated in the literature. The first, a linear combination of Dicke states, describes a family of target states with a potentially large number of parameters for which the entanglement can be accurately calculated using a combination of analytic and numerical methods. The second is a one parameter family of target states that are invariant under the interchange of any pair of even and/or odd qubits. 

Ultimately one would like to understand quantitatively the behaviour of  geometrical entanglement for large numbers of qubits. This is difficult to calculate in general so that previous work has necessarily focused on either small numbers of qubits or a high degree of symmetry to reduce the number of parameters. Our studies extend the calculable parameter space in an interesting and non-trivial way. 
%The affect of using complex coefficients had not been systematically studied in previous works. 
For a linear combination of  Dicke states we found evidence for  correlations between the distribution of coefficients and the entanglement, but there is also a great deal of interesting structure that requires further investigation.   \\[10pt]
\noindent
{\bf Acknowledgements:}  The authors are grateful to Dallas Clement for useful discussions. This research was supported by the Natural Sciences and Engineering Research Council (NSERC) of Canada and the Brandon University Research Committee (BURC) Research Award.

\end{document}